\documentclass[aps,prl,twocolumn,superscriptaddress,groupedaddress]{revtex4}

\usepackage{graphicx}  
\usepackage{dcolumn}   
\usepackage{bm}        
\usepackage{amssymb}   
\usepackage{soul, color}

\begin{document}
\title{\textit{Ab initio} Description of Bond-Breaking in Large Electric Fields}

\author{Michael Ashton}
\author{Arpit Mishra}
\author{J{\"o}rg Neugebauer}
\author{Christoph Freysoldt}
\email{c.freysoldt@mpie.de}
\affiliation{
Max-Planck-Institut f{\"u}r Eisenforschung,\\
Max-Planck-Stra{\ss}e 1, 40227 D{\"u}sseldorf, Germany
}


\begin{abstract}
  Strong (10$^{10}$~V/m) electric fields capable of inducing atomic bond-breaking represent a powerful
  tool for surface chemistry. However, their exact effects are difficult
  to predict due to a lack of suitable tools to probe their associated atomic-scale mechanisms.
  Here we introduce a generalized dipole correction for charged repeated-slab models that
  controls the electric field on both sides of the slab, thereby enabling direct theoretical
  treatment of field-induced bond-breaking events. As a prototype
  application, we consider field evaporation from a kinked W surface. We reveal two
  qualitatively different desorption mechanisms that can be selected by the magnitude of the
  applied field.

\end{abstract}

\maketitle

The breaking of an atomic bond is one of the most fundamental phenomena governing
materials transformation, reaction, and degradation. Phase changes, mechanical deformation,
chemical reactions, corrosion, and many other important processes can be understood in
very simple terms as a
systematic and often coordinated sequence of bond-breaking events. Probing and
controlling these processes
is therefore only possible with a clear understanding of the underlying effects that
stimulate bond-breaking.

\begin{figure}
  \includegraphics[width=\columnwidth]{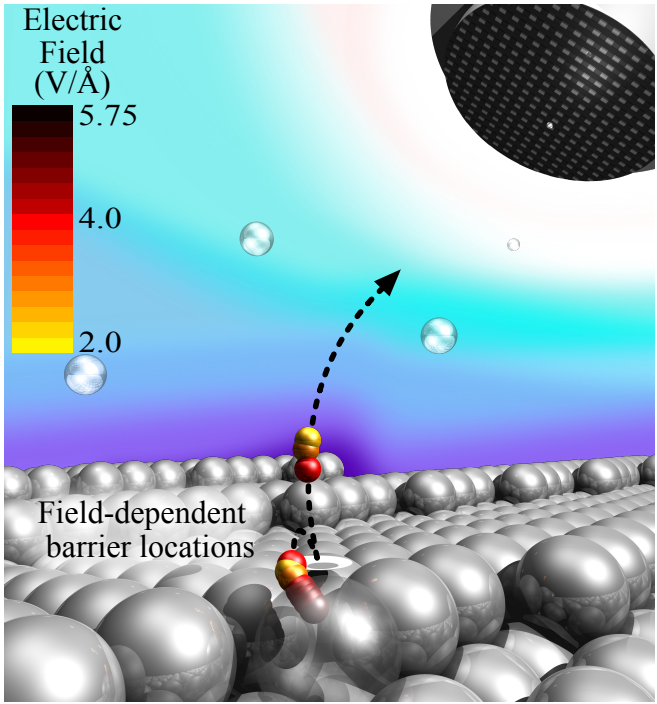}
  \caption{\label{fig:schematic} (Color online) a) A (10 8 1)
  tungsten surface to which varying electric fields are applied during the sequential
  removal of a surface atom to simulate its evaporation. The barrier
  configuration of the evaporating atom is shown as a function of applied field
  (small spheres, colored). At lower fields, the configuration for both of the
  barriers in the two-stage evaporation mechanism are shown.}
\end{figure}

Of the possible stimuli for bond-breaking, electric fields are among the most
ubiquitous. A local 10$^{10}$~V/m electric field is of the same magnitude as the intra-atomic
fields between electrons and
nuclei\cite{kreuzer2004physics} and is therefore perfectly capable of severing
atomic bonds. Because the field at a material's surface scales inversely with the local radius
of curvature, even
moderate voltages can be locally enhanced into fields of this magnitude anywhere that sharp
features exist, such as surface steps and kinks.\cite{yao2015effects}

This field enhancement enables atom probe tomography (APT), a microscopy
technique wherein nanosharp material samples are intentionally evaporated under strong fields. Ionized
atoms that evaporate from the sample's surface are later collected at a
counterelectrode (Figure~\ref{fig:schematic}).~\cite{muller1968atom,
blavette1993tomographic, kelly2012atom} After the evaporation,
the sample is computationally reconstructed by back-tracing each ion's trajectory
using its time-of-flight and detected location at the counterelectrode. The accuracy of
these projected trajectories, and consequently the accuracy of
the image reconstruction, depends on our understanding of the
mechanisms by which the original surface bonds were broken. 

Density functional
theory (DFT) calculations, which could enable a direct investigation
of evaporation mechanisms, are hindered by the challenge of applying a
finite electric field under periodic boundary conditions.\cite{fu1989external}
Here, we report on an efficient solution for this problem within a framework
of standard DFT calculations, and demonstrate its usefulness
for elucidating the complex evaporation mechanisms from prototypical kink sites on
high-index surfaces.

\begin{figure}
  \includegraphics[width=\columnwidth]{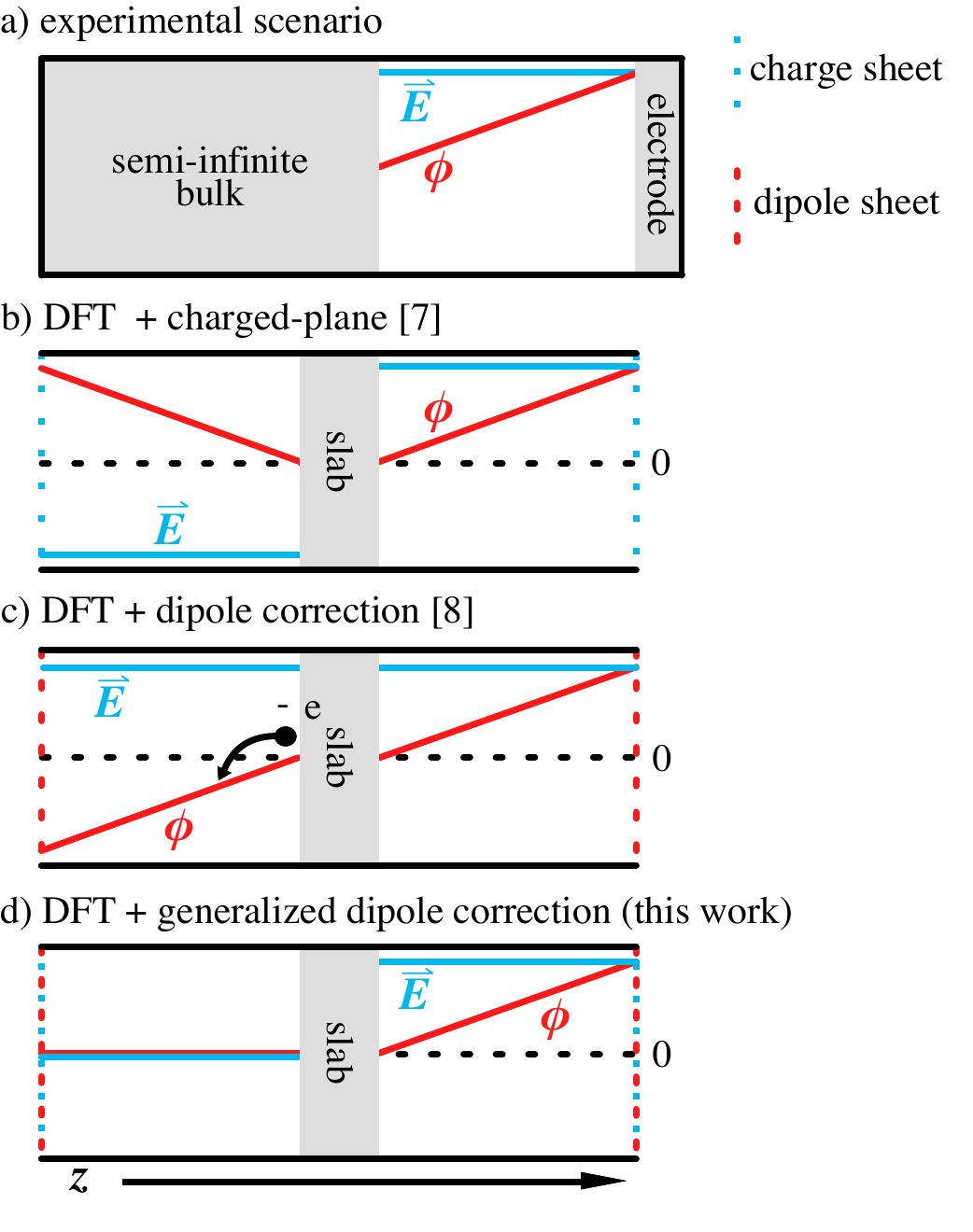}
  \caption{\label{fig:gdc} (Color online) Schematic representation of various
  approaches to model experimental electrostatic asymmetry (a) in DFT
  calculations: b) the charged-plane
  approach,\cite{sanchez2004field} with opposite fields on either side of the
  slab, c) the ``dipole correction",\cite{dipoleCorrection} with a constant field
  across the cell, and d) the generalized dipole correction (this work), with
  a field-exposed and field-free side. The resulting potential $\phi$ (red)
  and field $\vec{E}$ (blue) are shown for each scheme.}
\end{figure}

Under three-dimensional periodic boundary conditions, a surface must be
modeled as a two-dimensional slab; i.e. the model system will have a
surface on either side. This slab must be sufficiently thick and have enough
vacuum above and below it to prevent these two surfaces from artificially interacting,
and any applied electric field must be accounted for when it crosses the boundary
(Figure~\ref{fig:gdc}).

Previous DFT studies have accounted for the electric field by introducing
a sheet charge in the vacuum as the counterelectrode in an overall
charge-neutral setup, and enforcing symmetry along the $z$-axis to ensure that the
surface charge on either side of the slab is well-controlled
(Figure~\ref{fig:gdc} (b)).~\cite{fu1989external, lozovoi2001ab, sanchez2004field}
Using this approach, the
slab structure must not only be strictly symmetric, but also sufficiently
thick to converge the potential and Friedel oscillations beginning from either
side of the slab. The cell's vacuum region must also be enlarged to mitigate the
artificial Coulomb repulsion
between the two evaporating ions on either side of the slab. These constraints
reduce computational efficiency and restrict the approach's feasibility to simple cases,
such as an adatom evaporating from a flat surface.
Experimentally, however, the most relevant sites from which field
evaporation occurs are kink sites at the edges of terraces on the round
emitter surface where the local curvature induces strong field
enhancement.~\cite{gault2008estimation, yao2015effects}
In order to enable calculations for large surfaces that contain such
low-coordinated features, like the (10 8 1) tungsten surface shown in
Figure~\ref{fig:schematic}, the electric field's periodicity must be accounted
for in a more general way.

A starting point is the well-known ``dipole correction",\cite{dipoleCorrection} in which
an infinitesimally thin dipole sheet is added to create a discontinuous
potential jump in the vacuum region of a DFT cell (Figure~\ref{fig:gdc}(c)). The magnitude of this dipole
is chosen such that it exactly compensates the dipole of the slab, creating
a constant-field condition even for asymmetric slabs. This formalism
has been extended to introduce an additional finite field,\cite{neugebauer1993theory} but
above a critical field strength the vacuum potential is pulled below
the Fermi level. This results in the spurious transfer of electrons into
the vacuum (Figure~\ref{fig:gdc}(c)).\cite{kyritsakis2019atomistic}

They key concept we propose in this letter is to augment the dipole layer with a charged
monopole sheet. This charged layer acts as a
counterelectrode in the vacuum, creating a discontinuous jump not only in
the potential but also in the field. The result is a constant positive field
between the counterelectrode and one side of the slab while
the opposite side remains field-free (Figure~\ref{fig:gdc} (d)).
The dipole correction must then
compensate the dipole of the combined system (slab + counterelectrode).
This approach, which we term the generalized dipole correction (GDC), may be
conceptualized as a combination of the
counterelectrode and dipole correction schemes, as it introduces a symmetric charge compensation without
requiring that the slab itself be symmetric. This combination keeps the advantages
of the respective approaches, while eliminating their respective disadvantages.

The generalized dipole correction that must be added to the standard
electrostatic potential with periodic boundary conditions
for a slab with charge $Q$ reads
\begin{equation}
V^{\rm corr}(z) =\left\{\begin{array}{cll}
z\le z_0: & V_0 - \mathcal E^{\rm corr} \,z &-\, \frac{2\pi Q}{cA}\,z^2 \\[4pt]
z > z_0: & V_0 - \mathcal E^{\rm corr} (z-c)&-\, \frac{2\pi Q}{cA}(z-c)^2 
\end{array}\right.
\label{eq:Vcorr2}
\;.
\end{equation}
Here, $z_0$ is the cut position within the vacuum, $A$ and $c$ are the surface
area and height of the slab supercell, respectively, and $V_0$ is an
offset that brings the plane-averaged total potential $\overline V$ at
$z_0$ to a constant value (we use $\overline V(z_0) = -\mathcal E^{\rm top}z_0$ where $\mathcal E^{\rm top}$
is the electric field on the top side of the slab).
The $z^2$ term compensates for the implicit homogeneous background in
the periodic potential, while the correction field is
\begin{equation}
\mathcal E^{\rm corr} = 
                    \frac{2\pi Q}{A}
                    - \frac{4\pi\mu}{cA}
                    + \mathcal E^{\rm bottom}
\end{equation}
where $\mu$ is the charged slab's dipole moment with respect to $z_0$, and
$\mathcal E^{\rm bottom}$ the field on the bottom side, which is zero for the
field evaporation calculations performed in this work. %
If the above $\overline V(z_0)$-alignment is
used, a consistent electrostatic energy can be directly obtained from the total
potential and the total charge density $\rho^{\rm es}$ (including nuclear charges) as
\begin{equation}
E^{\rm es} = \frac 12 \int d^3\mathbf r~ V^{\rm es}(\mathbf r)\rho^{\rm es}(\mathbf r) 
   - \frac 12 \mu\, \mathcal E^{\rm bottom}
\end{equation}


The GDC has been implemented in our DFT code
SPHInX,\cite{SPHInX} allowing us to directly investigate evaporation mechanisms from
experimentally relevant surface sites using DFT.

To demonstrate the performance and applicability of the GDC approach, we consider
evaporation from a kinked tungsten surface. The field-dependence of the activation energy
for evaporation events in tungsten do not follow the behavior predicted by basic theoretical models,
\cite{kellogg1984measurement} which generally
assume an ideal straight-line departure of the ion from the surface. This discrepancy has prompted the
proposal of a number of nontrivial evaporation mechanisms, including possible out-of-sequence
evaporation,\cite{vurpillot2018simulation} a roll-up motion of
atoms onto neighboring step edges,\cite{wada1984thermally, schmidt1994binding,
waugh1976investigations, suchorski1996field}, or diffusion across the surface prior
to evaporation.\cite{sanchez2004field} A combination of several of these effects is
also possible.

To model field evaporation from this system, we model the kinked surface as six atomic layers
with a (10 8 1) surface normal, resulting in a 98 atom structure that has semi
close-packed (1 1 0) terraces with a single (0 0 1) step every 7 unit cells and
a (1 0 0) kink every 3 unit cells along the step.\cite{van1980new} A representation of this slab, which is given
15~\AA~of vacuum between its periodic images along the $z$-axis, is
shown in Figure~\ref{fig:schematic}. All DFT calculations are performed using
a local-density approximation (LDA)
functional with a 3$\times$3$\times$1 $k$-point mesh, a 20~Ry energy cutoff, and
0.1~eV Fermi smearing to allow partial electronic occupations. The LDA functional is
chosen based on its accurate reproduction of surface energies and work functions\cite{patra2017properties}
and the energy cutoff and k-point resolution give forces converged to within 10 meV/\AA
of a 7$\times$7$\times$1 $k$-point mesh and 30~Ry energy cutoff.

We geometrically optimize the surface at a number of field strengths, and then
pull an atom sitting at the kink site highlighted in Figure~\ref{fig:schematic}
from the surface by incrementing its $z$-coordinate
above its original position.\cite{sanchez2004field} The kink atom is chosen because
it is the surface atom with the fewest nearest neighbors and should thus be the
most weakly bound. Indeed, these are nearly always the first sites observed to
evaporate in experiments.\cite{gault2012atom}
At each incremental height the atom's $x$- and
$y$-coordinates, as well as the top three layers of the surface, are re-optimized
using a quasi-Newton algorithm based on the forces acting on each atom.
This enables us to observe how the coordinates of the evaporating atom and
the neighboring lattice change during the imposed evaporation event.

\begin{figure}
  \includegraphics[width=\columnwidth]{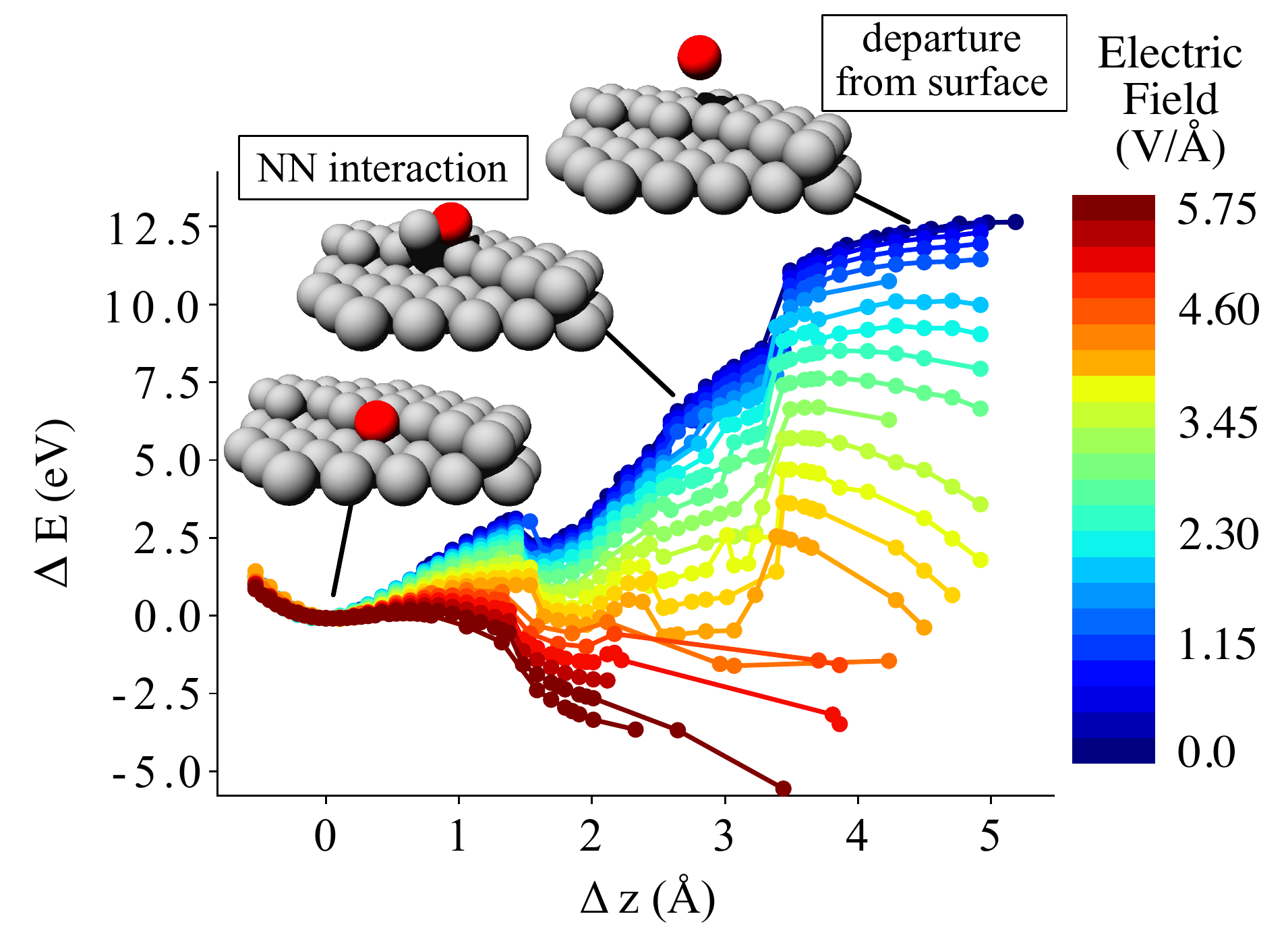}
  \caption{\label{fig:e_vs_d} (Color online) Energy as a function of field and
  height ($z$-coordinate) above its original position for a kink corner atom
  on a (10 8 1) tungsten surface. The structural insets exemplify the various
  stages of the evaporation mechanism (see text).}
\end{figure}

For a dense set of field strengths, we compute and plot the total energy as a function
of the evaporating atom's $z$-coordinate, as shown in
Figure~\ref{fig:e_vs_d}. The discontinuities in the potential energy curves in
Figure~\ref{fig:e_vs_d} are the result of discrete changes taking place along the
evaporation reaction coordinate. For distances where the evaporating atom is
very close to its original position ($<$1.5~\AA), the energy follows a smooth bond-stretching
trend. There is an abrupt shift to a new minimum between 1.5~-~2~\AA. At this height
the evaporating atom shifts laterally into the nearest hollow site atop
the neighboring (1 1 0) terrace. At lower heights this motion is sterically
prohibited by the neighboring step atoms.
The atom's evaporation then proceeds from this new
minimum in a manner very similar to an adatom on a flat (1 1 0) surface. The
second discontinuity
in the energy vs. distance profile occurs when the original atom begins to pull
up its nearest neighbor out of the step edge; at certain distances the two even
form a dimer above the surface. The persistent interaction between the
evaporating W atom and its nearest neighbor
is likely responsible for the substantial number of spatially correlated
co-evaporation events experimentally observed during tungsten
evaporation.\cite{muller2011some} At distances
sufficiently high above the surface ($>$3.3~\AA), the bond between these two atoms
becomes too weak to pull the neighboring atom up from its original position.

\begin{figure}
\includegraphics[width=8cm]{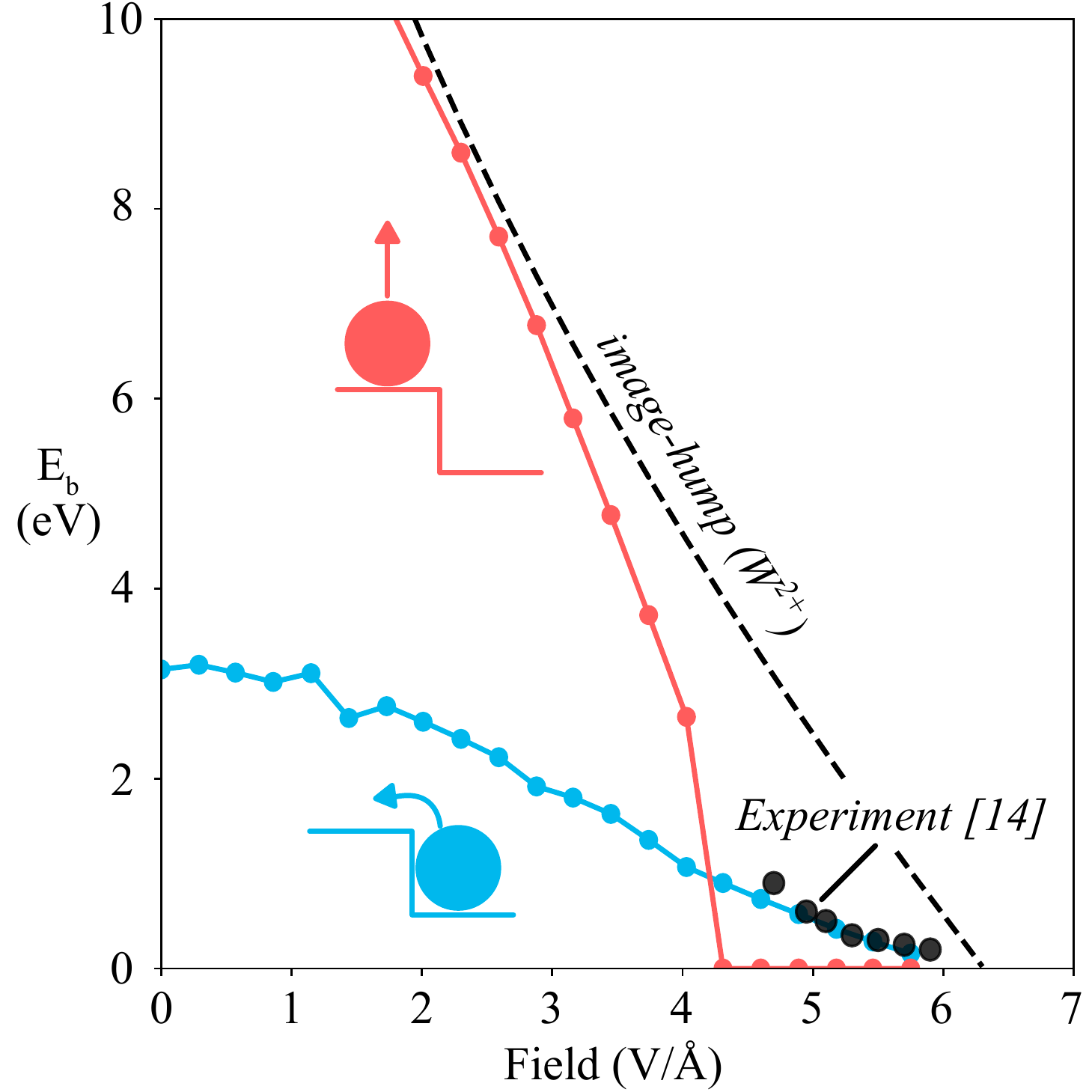}
\caption{\label{fig:barriers} (Color online) Energetic barriers as a function of applied
electric field for the W (10 8 1) kink atom. The barrier to the
initial rollover motion (blue) and the barrier to evaporation from
the adatom site (red) are plotted separately. Experimental data previously reported for
tungsten\cite{kellogg1984measurement} is shown in black. For comparison, the barrier
height for a W$^{2+}$ ion calculated using the image-hump model is also shown (dashed line).}
\end{figure}


Our calculations therefore reveal that the evaporation mechanism is effectively a two-stage
process: a rollover event followed by the actual departure from the surface. Each
of the two stages has its own respective energy barrier. The two barrier heights
vary quite differently as a function of field, as shown in
Figure~\ref{fig:barriers}. Since evaporation requires both mechanisms sequentially,
the effective barrier observed in experiments belongs to the mechanism with the
rate-limiting (i.e. highest) barrier. At low fields, the second step, in
which the atom is forced to ionize, has a much higher barrier.
However, the barrier in this step strongly decreases with increasing field. Due
to this rapid decrease, its barrier drops below that of the rollover stage's
barrier at a field strength close to 4.2~V/\AA. We note that above 4.2~V/\AA, the
atom still travels through the adatom site before evaporating, but
experiences no barrier after the initial rollover motion. From an energetic standpoint, therefore, the
mechanism effectively switches from two-stages to one-stage at high fields, although the
evaporating atom follows the same pathway as for low fields.

If the local field is below the critical value of
4.2~V/\AA, which can be the case on the shank of the emitter where the field is reduced, the
kink atom may be thermally
stimulated (e.g. by a nearby oncoming laser pulse) to roll over to the
adatom position but still not have enough energy to
evaporate. In this energetic trap atop the flat (1 1 0) terrace, the lateral hopping
barrier for an adatom is reduced from 0.9~eV to
around 0.7-0.8~eV by the
field.\cite{wang1982field, tsong1975direct}

The suppression of surface diffusion barriers by the field can be even more pronounced
depending on the material and nature of the diffusion mechanism.\cite{feibelman2001surface} Any
net displacement of the atom's position before it evaporates, whatever the mechanism, is
detrimental to the APT reconstruction's accuracy.\cite{plummer1968atomic, oberdorfer2018influence}
Therefore, experimental conditions should be chosen to avoid or mitigate diffusion from the
adatom trap.

Figure~\ref{fig:schematic} displays the location of the transition state (where
the evaporating atom experiences the barrier) as a function of field strength,
showing that the transition state exists very near the
atom's original location for fields above 4.2~V/\AA. For fields below 4.2~V/\AA,
transition states for both barriers are shown, including the second one above the adatom site.
The strong sensitivity of the location where the transition state exists renders
an often employed approximation - that the zero-field barrier configuration can
be used for all field strengths - invalid.

The \textit{ab initio}-calculated barriers can now be compared with those derived from existing models.
Historically, one of the most commonly used models to approximate field evaporation is the image-hump
model,\cite{muller1941abreissen, muller1956field, tsong1970field} which superimposes
the field potential and the image potential to determine the barrier height for
an ion leaving a flat surface. The main advantage of this model is its simplicity. The
only material-dependent parameters to enter the formula for barrier height are
the material's sublimation energy and relevant ionization energies, which in most cases
can simply be looked up. However, this and related models\cite{mckinstry1972examination}
have been proven to predict severely inaccurate
temperature-dependent evaporation fluxes.~\cite{wada1984thermally, kellogg1984measurement, tsong1988experimental,
gomer1994field, forbes1995field} In Figure~\ref{fig:barriers}, for example, we
compare the image-hump barrier for the evaporation of a W$^{2+}$ ion with
experimentamatcheslly determined W evaporation barriers.\cite{kellogg1984measurement}
The model nearly matches the extrapolated critical field of 6.2 V/\AA~ from our calculations,
but predicts unphysically high barriers for all other fields. Because APT
experiments are generally performed at fields below the critical field limit and
often use lasers to thermally stimulate evaporation,\cite{kellogg1980pulsed, gault2006design}
the model's predictions are invalid for exactly the experimentally most relevant
range of fields.

The failure of the image-hump model to describe the experimental data in
Figure~\ref{fig:barriers} has led to the recent proposal of several new analytical
models to calculate evaporation barriers.\cite{miller2009atom, gault2012atom}
One of the most commonly accepted is the so-called ``charge-draining'' model, in which
the evaporating atom is considered to continuously donate
charge to the slab and gradually ionize as it departs the surface. Previous DFT calculations on
charged Al (1 1 1) adatoms support this nature of charge transfer.\cite{sanchez2004field}
Due to the continuous ionization, this model yields an evaporation barrier that decays
linearly as the field is increased. Unfortunately, these models contain effective parameters
that must be obtained from external sources. In practice, the slope of the decay is generally
fit empirically to available experimental data. However, since this slope depends
directly on the shape and size of the barrier encountered by the evaporating atom,\cite{wang1990kinetic, miller2009atom}
potential energy paths calculated using DFT with the GDC approach, as in
Figure~\ref{fig:e_vs_d}, are a reliable route to provide quantitative
accuracy to these more conceptually sophisticated analytical models.

We conclude that conventional APT experiments in tungsten automatically
probe the rollover response, which is ultimately detrimental to their 3D spatial
resolution. The rollover response can be understood as a
competition between the force of the evaporating atom's nearest neighbor bonds
and the force of the field tugging on the ion. As a result, softer metals with
weaker surface bonds are expected to exhibit a less pronounced version of this
effect than what is observed here for tungsten.

The two-stage rollover evaporation mechanism provides a natural explanation for the experimentally
observed evaporation barrier versus applied field in Figure~\ref{fig:barriers}. It clarifies that at very high fields, experimental evaporation
events are dominated by a thermally-activatable rollover barrier. Of course, observations from APT experiments also depend
on several phenomena which occur at length and time scales inaccessible to DFT, including mesoscopic field and
temperature gradients. The atomic-scale evaporation mechanism is therefore an important piece
of the overall theory of field evaporation in APT, which requires
considerations beyond DFT to account for these larger-scale phenomena.

The evidence provided in this study for a field-dependent, tunable evaporation mechanism
is essential for accurately controlling and interpreting APT and field ion microscopy
experiments on metallic systems. The generalized dipole correction developed here provides
a computationally efficient and easily implementable approach to model the effect of strong electric fields in DFT
calculations. The correction is universally applicable to other material systems
in order to understand bond-breaking mechanisms in more complex materials systems,
e.g. aqueous corrosion systems or catalytic surfaces. Using this technique to directly
probe the response of materials and chemical reactions, such as bond-breaking,
in extreme electric fields will provide a new tool to
guide the interpretation and design of new experiments and applications.

The authors gratefully acknowledge funding via BiGmax, the Max Planck Society’s
Research Network on Big-Data-Driven Materials-Science.

\bibliography{References}

\end{document}